# Improved Currents for Heavy Quarks

Andreas S. Kronfeld[a]

[a]Theoretical Physics Group, Fermi National Accelerator Laboratory,
P.O. Box 500, Batavia, Illinois 60510-0500, U.S.A.

We discuss lattice artifacts for matrix elements of hadrons containing one or more heavy quark. In particular, we analyze interrelations between lattice artifacts and the $1/m_Q$ expansion. The implications for calculations of heavy-light decay constants and of semi-leptonic form factors are discussed.

## 1. INTRODUCTION

During the past several years, we have discussed how to make sense of Monte Carlo computations of Wilson fermions, even when the fermion mass becomes large [1–4]. (By "Wilson fermions" I mean a four-component field and Wilson's solution [5] of the doubling problem.) Because the numerical data never satisfy $m_Q a \ll 1$ and often have $\Lambda_{\rm QCD}/m_Q \sim \frac{1}{2}$, we choose not to assume either limit when calculating couplings of the improved action, or renormalization of composite operators. If warranted, these limits can be recovered afterwards.

This report concentrates on external operators, whose matrix elements are needed to study electroweak properties of heavy-light hadrons. For comprehensive exposition with lattice terminology, see ref. [6]. To complement that article, this one often uses the language of heavy-quark effective theory instead. Nevertheless, eqs. (7) and (13) are valid for all $m_Q a$ and $m_Q/\Lambda_{\rm QCD}$ [6].

The completely renormalized (i.e. cutoff-free) action is

$$\mathcal{S}(m_Q/\Lambda_{\rm QCD}) = \sum_n c_n(m_Q a, g_0^2) S_n, \quad (1)$$

where the sum runs over all interactions $S_n$ with the desired field content and symmetries. Similarly, the completely renormalized operator is

$$\mathcal{O} = Z_{\mathcal{O}}(\{m_0 a\}, g_0^2) \sum_n C_n(\{m_0 a\}, g_0^2) O_n, \quad (2)$$

where the $O_n$ are all operators with the correct quantum numbers. In practice one truncates the sums and computes couplings $c_n$, normalization factors $Z_{\mathcal{O}}$, and coefficients $C_n$ only approximately. We match on-shell amplitudes computed via tadpole-improved perturbation theory in $g_0^2$. At every fixed order in $g_0^2$, however, we match to *all* orders in $m_0 a$. (Higher-order corrections $C_n^{[l]}$ are mass dependent too, as clearly stated in ref. [3] and explicitly demonstrated in ref. [4].)

Here we truncate eqs. (1) and (2) at tree level ($l = 0$), dim$S_n \leq 5$, and dim$O_n \leq 1 + $dim$\mathcal{O}$. Hence, our construction, like any other, yields an action $S = \mathcal{S} + \delta S$ and operators $O = \mathcal{O} + \delta O$ that are imperfect. To make sense of a numerical calculation with $S$ and $O$, one must understand the errors stemming from $\delta S$ and $\delta O$.

## 2. HEAVY-QUARK SYMMETRY

Treating the gauge field semi-classically, the Hamiltonian (logarithm of the transfer matrix) can be written [2,6]

$$\hat{H} = \hat{\bar{\Psi}}_{\rm NR} \left[ M_1 + \gamma_0 A_0 - \frac{\boldsymbol{D}^2}{2M_2} - \frac{i\boldsymbol{\Sigma}\cdot\boldsymbol{B}}{2M_B} \right] \hat{\Psi}_{\rm NR}. \quad (3)$$

The non-relativistic field $\hat{\Psi}_{\rm NR}(x)$ is related to $\psi(x)$, which appears in the action, by a Foldy-Wouthuysen-Tani transformation and a certain normalization factor [2,6]. Neglecting terms of order $(\boldsymbol{p}a)^2$ and higher,

$$\Psi_{\rm NR}(x) = \exp(\tfrac{1}{2}M_1 a)\left[1 + a\xi_{\rm NR}\boldsymbol{\gamma}\cdot\boldsymbol{D}\right]\psi(x). \quad (4)$$

The "masses" $M_1$, $M_2$, and $M_B$ depend the couplings $\zeta$, $m_0 a$ and $c_B$, defined in ref. [2,6].

The Hamiltonian obeys heavy-quark symmetries for infinite mass. In that limit the flavor-and spin-blind Coulomb term generates the only dynamics. The corrections to the flavor symmetry are correct only if $M_2 = m_Q$ (so tune the bare mass $m_0 a$), and the corrections to the spin symmetry are correct only if $M_B = m_Q$ (so tune the coefficient of the clover term [7] $c_B$).

From the heavy-quark effective theory [8], one knows to construct currents, etc., not with the non-relativistic field, but with

$$\Psi_I(x) = [1 - \boldsymbol{\gamma} \cdot \boldsymbol{D}/(2m_Q)] \Psi_{NR}(x). \quad (5)$$

Combining eq. (4) and (5)

$$\Psi_I(x) = \exp(\tfrac{1}{2} M_1 a) \left[1 + a d_1 \boldsymbol{\gamma} \cdot \boldsymbol{D}\right] \psi(x), \quad (6)$$

where

$$d_1 = \xi_{NR} - \frac{1}{2m_Q a} = \frac{\zeta(1 + m_0 a)}{m_0 a (2 + m_0 a)} - \frac{1}{2M_2 a}. \quad (7)$$

The last equality assume $M_2 = m_Q$ to obtain the correct amount of heavy-quark flavor symmetry breaking. With $\Psi_I(x)$ operators take the simple form

$$J_\Gamma^{fg}(x) = \bar{\Psi}_I^f(x) \Gamma \Psi_I^g(x). \quad (8)$$

## 3. IMPROVED OPERATORS

Consider the construction in eq. (2). There is a redundancy because, for example, one can use either local or point-split operators. Consider a modification of the local current

$$V_\mu^L = Z_{V_\mu^L} \bar{\psi} \left[\gamma_\mu + aC \left(\boldsymbol{\gamma} \cdot \overleftarrow{\boldsymbol{D}} \gamma_\mu + \gamma_\mu \boldsymbol{\gamma} \cdot \boldsymbol{D}\right)\right] \psi. \quad (9)$$

The role of the dimension-4 term will emerge shortly. Table 1 contains matrix elements of $V_\mu^L$ and the conserved gauge current $V_\mu^G$. ($V_\mu^G$ differs from the Noether current if the action has a $\sigma \cdot F$ term.) All three nonzero matrix elements are correct for the local current, if $Z_{V_0^L} = Z_{V_i^L} = e^{M_1 a}$ and if $C$ is tuned so that $M_V = M_2$.

The "form-factor" matrix elements of the gauge current would require $Z_{V_0^G} = Z_{V_i^G} = 1$, but the "decay-constant" matrix element would require a different $Z_{V_i^G}$ for $M_1 \neq 0$. One can obtain all three matrix elements correctly only by taking $Z_{V_i^G} \neq Z_{V_0^G}$ and adding to $V_i^G$ a term $aC'\bar{\psi} D_i \psi$, as in $V_i^L$. But the new term spoils the attraction of the gauge current, namely that current conservation determines its normalization exactly. Hence, the gauge current seems to be even less practicable than the local current.

To tune $M_V = M_2$ at tree level one must choose $C = d_1$. Indeed, with $Z_{V^L} = e^{M_1 a}$, eq. (9) is the same as the construction of eqs. (5)–(8). One finds that $d_1 \approx (1 - \zeta_0'')m_0 a/4$ for small mass, when the only O($a$) correction is in $Z_{V^L}$. On the other hand, one finds that $d_1 = (2m_Q a)^{-1}$ in the infinite mass limit. Without the improvement, the operator is incorrect at O($1/m_Q$).

## 4. ERRORS ANALYSIS FOR $\phi_M$

Improved Wilson fermions obey heavy-quark symmetry, even when the quantities $c_B$ and $d_1$ are maladjusted. Thus, one can exploit the machinery of heavy-quark effective theory to estimate the associated errors. We shall illustrate the method with the decay constant $\phi_M = \sqrt{m_M} f_M$, where $M = (P, V)$ denotes pseudoscalar and vector heavy-light mesons. Similar results hold for semi-leptonic form factors.

According to heavy-quark effective theory

$$\phi_M = \phi_\infty \left(1 - \frac{A_M}{m_M} + \cdots\right), \quad (10)$$

where $\phi_\infty$ is the matrix element, evaluated at infinite heavy-quark mass. The $1/m_M$ correction has three distinct contribution,

$$A_M = G_1 + 6h_M G_2 + \tfrac{1}{2} h_M \bar{\Lambda}, \quad (11)$$

from the heavy-quark kinetic energy, chromomagnetic interaction, and $1/m_Q$ expansion of the weak current (eqs. (5) and (8)) [8]. The helicity factors are $h_P = 1$ and $h_V = -1/3$. Each contribution to $A_M$ can be related to a matrix element in the infinite mass theory. ($G_i > 0$ and $\bar{\Lambda} = m_M - m_Q$.)

With eqs. (3) and (9) the slope becomes

$$\begin{aligned} A_M^{\text{lat}} &= \frac{m_Q}{M_2} G_1 + h_M \left(6 \frac{m_Q}{M_B} G_2 + \tfrac{1}{2} X \bar{\Lambda}\right), \\ &= G_1 + h_M \left(6 \frac{M_2}{M_B} G_2 + \tfrac{1}{2} X \bar{\Lambda}\right), \end{aligned} \quad (12)$$

where the second line assumes $m_Q = M_2$. At tree

Table 1
Matrix elements of local and conserved vector currents. The quantity $M_V$ depends on the coefficient $C$ in eq. (9). Some common factors are suppressed, but the third column gives the bottom line.

| matrix element | local | conserved | should be |
| --- | --- | --- | --- |
| $\langle Q(\boldsymbol{p})\|V_0\|Q(\boldsymbol{p})\rangle$ | $Z_{V_0^{\rm L}} e^{-M_1 a}$ | $Z_{V_0^{\rm G}}$ | 1 |
| $\langle Q(\boldsymbol{p})\|V_i\|Q(\boldsymbol{p})\rangle$ | $Z_{V_i^{\rm L}} e^{-M_1 a} p_i/M_V$ | $Z_{V_i^{\rm G}} p_i/M_2$ | $p_i/m_Q$ |
| $\langle 0\|V_0\|Q(\boldsymbol{p})\bar{Q}(-\boldsymbol{p})\rangle$ | 0 | 0 | 0 |
| $\langle 0\|V_i\|Q(\boldsymbol{p})\bar{Q}(-\boldsymbol{p})\rangle$ | $Z_{V_i^{\rm L}} e^{-M_1 a}$ | $Z_{V_i^{\rm G}} \zeta[1 + \tfrac{1}{2} c_E \sinh(2M_1 a)] e^{-M_1 a}$ | 1 |

level

$$X = 2 M_2 a \left( \frac{\zeta(1+m_0 a)}{m_0 a(2+m_0 a)} - C \right). \qquad (13)$$

Finally, $G_i^{\rm lat} - G_i$ and $\bar{\Lambda}^{\rm lat} - \bar{\Lambda}$ are suppressed by a power of $a$, so I neglect them here.

If $c_B = 0$ (so $M_B \neq M_2$), $M_2/M_B \to 0$ as $m_0 a \to \infty$; hence the Wilson action *under*estimates the chromomagnetic contribution to $A_M$. If $C = 0$ (instead of $d_1$, eq. (7)), $X \to 2\zeta$ as $m_0 a \to \infty$; hence the usual local current *over*estimates the current contribution to $A_M$. In both cases the relative error in $A_M$ is $O(a^0)$. In contrast, maladjustments in the heavy-quark $c_B$ and $d_1$ produce *no* error in $\phi_\infty$.

On the other hand, if $c_B(m_0 a, g_0^2)$ and $d_1(m_0 a, g_0^2)$ can be computed reliably in tadpole-improved perturbation theory, (our version of) the improvement program reduces $A_M - A_M^{\rm lat}$ to $O(g_0^{2l})$. Nonperturbative adjustments of $M_B$ and $M_V$ could reduce the difference further.

## 5. CONCLUSIONS

In our analysis of the action [2,3,6], we found that there were only two large effects associated with the large mass limit of actions for Wilson fermions. One is the distinction between $M_2$ and $M_1$, and the other is the field normalization factor $e^{M_1 a/2}$ [1–3,6]. Eqs. (6), (8) and sect. 3 show how the mass dependence of the unobservable field normalization filters down to currents and other multi-quark operators.

With mass-dependent improvement the remaining lattice artifacts are

$$\delta S \sim c_n^{[l]}(m_0 a) \times (a\boldsymbol{p})^{\dim S_n - 4},$$
$$\delta O \sim C_n^{[l]}(m_0 a) \times (a\boldsymbol{p})^{\dim O_n - \dim O}, \qquad (14)$$

where $\boldsymbol{p}$ is a dynamical scale. Experience, both perturbative [4] and nonperturbative [9], indicates that the residual mass dependence interpolates smoothly from small-mass to static results, with a "knee" at $M_2$ between 0 and 5. In particular, the $c_n$ and $C_n$ do not grow uncontrolledly.

## ACKNOWLEDGMENTS

This paper describes work done with Aida El-Khadra and Paul Mackenzie. I would like to thank George Hockney, Bart Mertens, Tetsuya Onogi, and Jim Simone for collaboration on related projects. Fermilab is operated by Universities Research Association, Inc., under contract DE-AC02-76CH03000 with the U.S. Department of Energy.